\documentclass[twocolumn,tighten,times]{aastex62}
\usepackage{color}

\graphicspath{{./}{figures/}}

\received{\today}
\revised{tomorrow}
\accepted{the day after tomorrow}

\submitjournal{ApJ}

\shorttitle{Alignment of Dwarfs in Fornax}
\shortauthors{Rong et al.}

\begin{document}

\title{The Next Generation Fornax Survey (NGFS): VI.~The Alignment of Dwarf Galaxies in The Fornax Cluster}

\correspondingauthor{Yu Rong}
\email{rongyuastrophysics@gmail.com}

\author[0000-0002-2204-6558]{Yu Rong}
\altaffiliation{FONDECYT Postdoctoral Fellow}
\affiliation{Institute of Astrophysics, Pontificia Universidad Cat\'olica de Chile, Av.~Vicu\~na Mackenna 4860, 7820436 Macul, Santiago, Chile}

\author[0000-0003-0350-7061]{Thomas H.~Puzia}
\affiliation{Institute of Astrophysics, Pontificia Universidad Cat\'olica de Chile, Av.~Vicu\~na Mackenna 4860, 7820436 Macul, Santiago, Chile}

\author[0000-0001-8654-0101]{Paul~Eigenthaler}
\affiliation{Institute of Astrophysics, Pontificia Universidad Cat\'olica de Chile, Av.~Vicu\~na Mackenna 4860, 7820436 Macul, Santiago, Chile}

\author[0000-0001-7966-7606]{Yasna Ordenes-Brice\~no}
\altaffiliation{PUC-HD Graduate Student Exchange Fellow}
\affiliation{Institute of Astrophysics, Pontificia Universidad Cat\'olica de Chile, Av.~Vicu\~na Mackenna 4860, 7820436 Macul, Santiago, Chile}
\affiliation{Astronomisches Rechen-Institut, Zentrum f\"ur Astronomie der Universit\"at Heidelberg, M\"onchhofstra{\ss}e 12-14, D-69120 Heidelberg, Germany}

\author[0000-0003-3009-4928]{Matthew A.~Taylor}
\altaffiliation{Gemini Science Fellow}
\affiliation{Gemini Observatory, Northern Operations Center, 670 North A'ohoku Place, Hilo, HI 96720, USA}

\author[0000-0003-1743-0456]{Roberto~P.~Mu\~noz}
\affiliation{Institute of Astrophysics, Pontificia Universidad Cat\'olica de Chile, Av.~Vicu\~na Mackenna 4860, 7820436 Macul, Santiago, Chile}

\author[0000-0003-1632-2541]{Hongxin Zhang}
\affiliation{CAS Key Laboratory for Research in Galaxies and Cosmology, Department of Astronomy, University of Science and Technology of China, China}

\author[0000-0002-8835-0739]{Gaspar Galaz}
\affiliation{Institute of Astrophysics, Pontificia Universidad Cat\'olica de Chile, Av.~Vicu\~na Mackenna 4860, 7820436 Macul, Santiago, Chile}

\author[0000-0002-5897-7813]{Karla~Alamo-Mart\'inez}
\altaffiliation{FONDECYT Postdoctoral Fellow}
\affiliation{Institute of Astrophysics, Pontificia Universidad Cat\'olica de Chile, Av.~Vicu\~na Mackenna 4860, 7820436 Macul, Santiago, Chile}

\author[0000-0002-3004-4317]{Karen X.~Ribbeck}
\affiliation{Institute of Astrophysics, Pontificia Universidad Cat\'olica de Chile, Av.~Vicu\~na Mackenna 4860, 7820436 Macul, Santiago, Chile}

\author[0000-0002-1891-3794]{Eva K.\ Grebel}
\affiliation{Astronomisches Rechen-Institut, Zentrum f\"ur Astronomie der Universit\"at Heidelberg, M\"onchhofstra{\ss}e 12-14, D-69120 Heidelberg, Germany}

\author[0000-0002-5322-9189]{Sim\'on~\'Angel}
\affiliation{Institute of Astrophysics, Pontificia Universidad Cat\'olica de Chile, Av.~Vicu\~na Mackenna 4860, 7820436 Macul, Santiago, Chile}

\author[0000-0003-1184-8114]{Patrick C{\^o}t{\'e}}
\affiliation{NRC Herzberg Astronomy and Astrophysics, 5071 West Saanich Road, Victoria, BC V9E 2E7, Canada}

\author[0000-0002-8224-1128]{Laura Ferrarese}
\affiliation{NRC Herzberg Astronomy and Astrophysics, 5071 West Saanich Road, Victoria, BC V9E 2E7, Canada}

\author[0000-0002-2363-5522]{Michael Hilker}
\affiliation{European Southern Observatory, Karl-Schwarzchild-Str. 2, D-85748 Garching, Germany}


\author[0000-0003-4197-4621]{Steffen Mieske}
\affiliation{European Southern Observatory, 3107 Alonso de C\'ordova, Vitacura, Santiago}

\author[0000-0002-5665-376X]{Bryan W.~Miller}
\affiliation{Gemini Observatory, South Operations Center, Casilla 603, La Serena, Chile}

\author[0000-0003-4945-0056]{Ruben S\'anchez-Janssen}
\affiliation{STFC UK Astronomy Technology Centre, Royal Observatory, Blackford Hill, Edinburgh, EH9 3HJ, UK}

\author[0000-0002-2368-6469]{Evelyn J. Johnston}
\altaffiliation{FONDECYT Postdoctoral Fellow}
\affiliation{Institute of Astrophysics, Pontificia Universidad Cat\'olica de Chile, Av.~Vicu\~na Mackenna 4860, 7820436 Macul, Santiago, Chile}

\begin{abstract}
	
Using the photometric data from the Next Generation Fornax Survey, we find a significant radial alignment signal among the Fornax dwarf galaxies.
For the first time, we report that the radial alignment signal of nucleated dwarfs is stronger than that of non-nucleated ones at 2.4$\sigma$ confidence level, and the dwarfs located in the outer region ($R>R_{\rm{vir}}/3$; $R_{\rm{vir}}$ is the Fornax virial radius) show slightly stronger radial alignment signal than those in the inner region ($R<R_{\rm{vir}}/3$) at $1.5\sigma$ level. We also find that the significance of radial alignment signal is independent of the luminosities or sizes of the dwarfs.

\end{abstract}



\section{Introduction} \label{sec:intro}

The major axes of massive galaxies usually align with the gravitational potential of their host large-scale structure, e.g., galaxy clusters/groups, filaments, sheets, etc., and thus provide clues to the co-evolution of the galaxies and their parent large-scale structures. For the member massive galaxies in clusters/groups, galaxy alignments have two primary sub-classifications, i.e., the radial alignment, and primordial alignment (also called direct alignment). The radial alignment, as the production of the tidal force of the parent cluster (e.g., Ciotti \& Dutta 1994; Pereira, Bryan \& Gill 2008; Usami \& Fujimoto 1997; Rong et al. 2015a), means that the major axes of galaxies tend to point to the cluster center (i.e., the radial angles $\phi$ between the major axes of galaxies and radial directions 
tend to be zero; see also Fig.~1 in Rong et al. 2015b), and therefore, is tightly related to the locations of the member galaxies in the cluster and cluster dark matter profile such as the concentration and scale radius (e.g., Rong et al. 2015a). The primordial alignment is referred to as the alignment between the major axes of galaxies and the elongations of the host clusters. Since the axis of the central brightest cluster galaxy (BCG) or cD galaxy in a cluster of galaxies strongly coincides with the elongation of the parent cluster (e.g., West 1994; Fuller et al. 1999; Struble 1990), the primordial alignment of the cluster member galaxies is also referred to as the alignment between the major axes of these galaxies and BCG (i.e., the angles $\theta$ between the major axes of member galaxies and BCG tend to be zero; see also Fig.~1 in Rong et al. 2015b). Primordial alignment has been used as a probe of the dynamical state of clusters (e.g., Plionis et al. 2003; Rong et al. 2015b) and surrounding large-scale filaments (Rong et al. 2016), and can be possibly used to constrain galaxy formation models and their interaction with large-scale structure (Hung et al. 2010). Galaxy alignments are also important contaminations in weak lensing measurements \citep{Hirata04}. The ellipticity of a galaxy can be subject to physical effects that stretch it and orient it in preferential directions with respect to large-scale structure \citep{Troxel14}, which can mimic the coherent galaxy alignments of gravitational lensing.


However, we still lack the alignment information of dwarf galaxies in clusters/groups due to the their extremely small sizes and dim surface brightness, and thus their discoveries are a challenge to many photometry surveys. 
The deep multi-wavelength Next Generation Fornax Survey (NGFS; Mu\~noz et al. 2015), covering the entire Fornax galaxy cluster out to its virial radius ($R_{\rm{vir}}\simeq 1.4$~Mpc, Drinkwater et al. 2001), provides the $u', g'$, and $i'$-band photometries (reaching point-source detections with S/N$\sim5$ at 26.5, 26.1, and 25.3~mag, respectively) obtained with the Dark Energy Camera (DECam; Flaugher et al. 2015) mounted on the 4m Blanco telescope at Cerro Tololo Interamerican Observatory (CTIO) and $J$ and $K_s$-band photometry from VIRCam mounted on the 3.7~m VISTA at ESO's Paranal Observatory (Sutherland et al. 2015). NGFS encompasses the unprecedented accuracy to study the properties of the extremely-faint and low-mass objects ($\sim 10^6\ M_{\odot}$) in the nearby cluster, including the stellar masses and ages of the nuclear star clusters (NSCs) in the nucleated dwarf galaxies \citep{Yasna18b}, color distributions of NSCs and dwarf spheroids \citep{Eigenthaler18}, dwarf clustering \citep{Yasna18a}, and in particular, morphologies of dwarfs. Further, since the morphologies of dwarfs are relevant to their properties and environments, e.g., the presence of NSCs, galactic luminosities/stellar masses, and their locations in galaxy clusters/groups \citep[e.g.,][]{Sanchez-Janssen19,Sanchez-Janssen10,Roychowdhury13}, it is reasonable to compare the alignment signals of the NGFS dwarfs with the different properties.



\section{Alignments of the Fornax dwarfs}\label{sec:2}


\begin{table*} \footnotesize
	\centering
\begin{tabular}{@{}lccccccc@{}}
\hline
\hline
             & \multicolumn{3}{c}{Distribution of $\phi$} & \multicolumn{3}{c}{Distribution of $\theta$} \\
Dwarf sample & $\alpha$ & $p\rm{(K-S)}$ & $p\rm{(K)}$ & $\alpha$ & $p\rm{(K-S)}$ & $p\rm{(K)}$ \\
\hline
All &  $0.71\pm0.06$ & $10^{-5}$ & $10^{-6}$  & $0.52\pm0.05$  & 0.82 &  0.52  \\
Nucleated & $0.96\pm0.12$  & $10^{-5}$ &  $10^{-5}$  & $0.56\pm0.10$  &  0.72 & 0.81  \\
Non-nucleated & $0.63\pm0.07$  & $4\times10^{-2}$ &  $5\times10^{-2}$  &  $0.50\pm0.05$ & 0.41  &  0.34 \\
$M_{i'}<-12.5$~mag & $0.69\pm0.10$  & $5\times10^{-3}$ &  $5\times10^{-3}$  &  $0.59\pm0.09$ &  0.67 & 0.30  \\
$M_{i'}>-12.5$~mag & $0.73\pm0.08$  & $3\times10^{-4}$ &  $2\times10^{-3}$  & $0.47\pm0.06$  & 0.80 &  0.62  \\
$R<R_{\rm{vir}}/3$ & $0.63\pm0.09$  & $2\times10^{-2}$ &  $6\times10^{-2}$  & $0.53\pm0.07$  & 0.69 &  0.28  \\
$R>R_{\rm{vir}}/3$ & $0.81\pm0.08$  & $10^{-4}$ &  $4\times10^{-4}$  & $0.52\pm0.07$  & 0.83 &  0.88  \\
$r_{\rm{e}}>0.6$~kpc & $0.73\pm0.09$  & $10^{-3}$ &  $4\times10^{-3}$  &  $0.61\pm0.08$ & 0.41 &  0.19  \\
$r_{\rm{e}}<0.6$~kpc &  $0.70\pm0.09$ & $5\times10^{-3}$ & $10^{-2}$   & $0.45\pm0.06$  &  0.51 &  0.72  \\
\hline
\hline
\end{tabular}
\caption{The $\alpha$ and $p$ values returned from the K-S test ($p\rm{(K-S)}$) and Kuiper test ($p\rm{(K)}$) for the distributions of $\phi$ and $\theta$ of the different dwarf samples.}
\label{table}
\end{table*}

\begin{figure*}[!]
\centering
\includegraphics[scale=0.14]{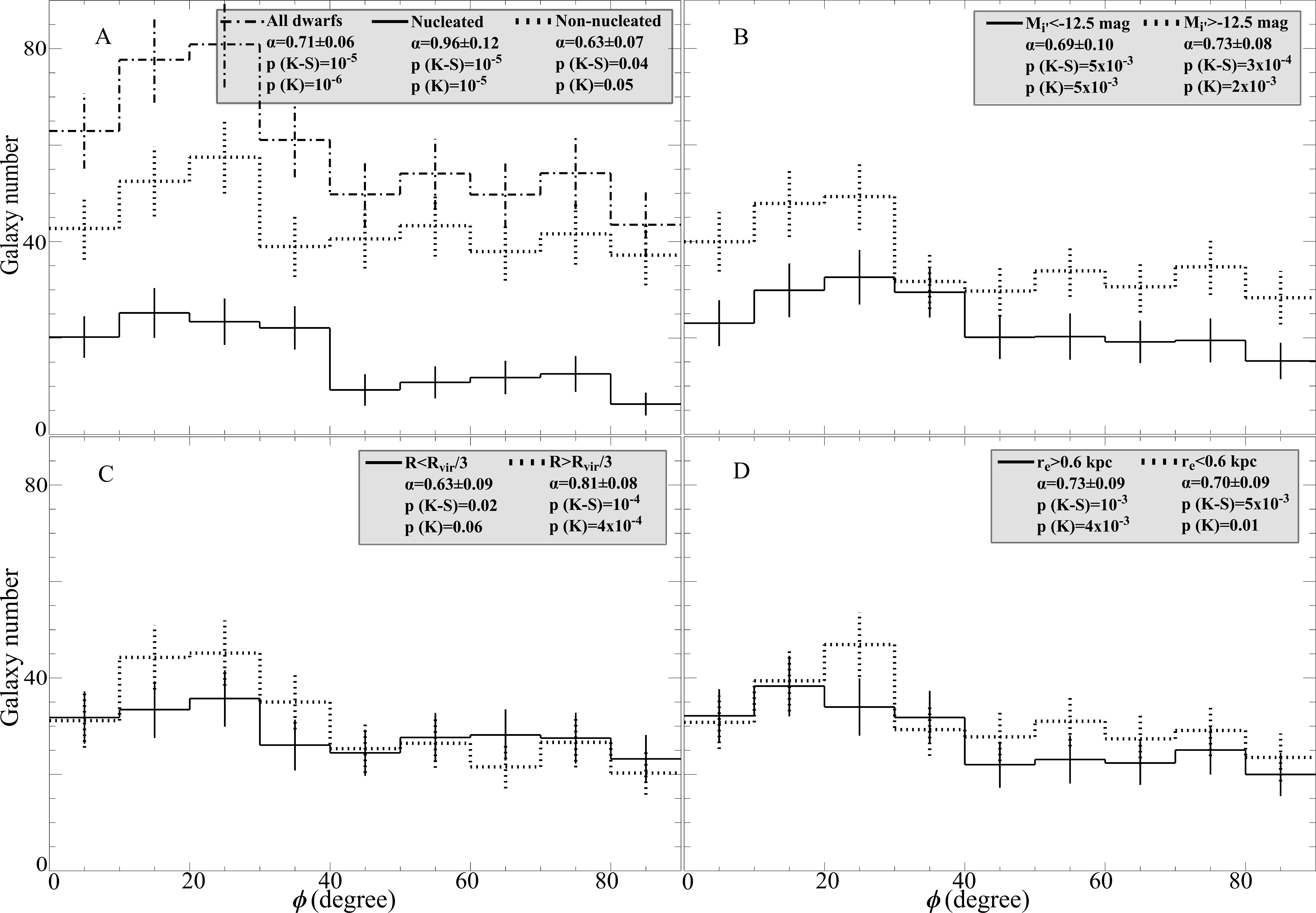}
\caption{\small Panel~A: RADs for all of the dwarfs (dash-dotted), nucleated (solid), and non-nucleated (dotted) dwarfs in the Fornax cluster. Panel~B: RADs for the bright ($M_{i'}<-12.5$~mag; solid) and faint ($M_{i'}<-12.5$~mag; dotted) dwarfs. Panel~C: RADs for the inner-region ($R<R_{\rm{vir}}/3$; solid) and outer-region ($R>R_{\rm{vir}}/3$; dotted) dwarf samples. Panel~D: RADs for the large ($r_{\rm{e}}>0.6$~kpc; solid) and small ($r_{\rm{e}}<0.6$~kpc; dotted) dwarfs. Hereafter, the error bars in the angle bins of each distribution are always estimated with the bootstrap methodology. $p$(K-S) and $p$(K) show the $p$ values from the K-S test and Kuiper test, respectively.}
\label{RA_final}
\end{figure*}

\begin{figure*}[!]
\centering
\includegraphics[scale=0.14]{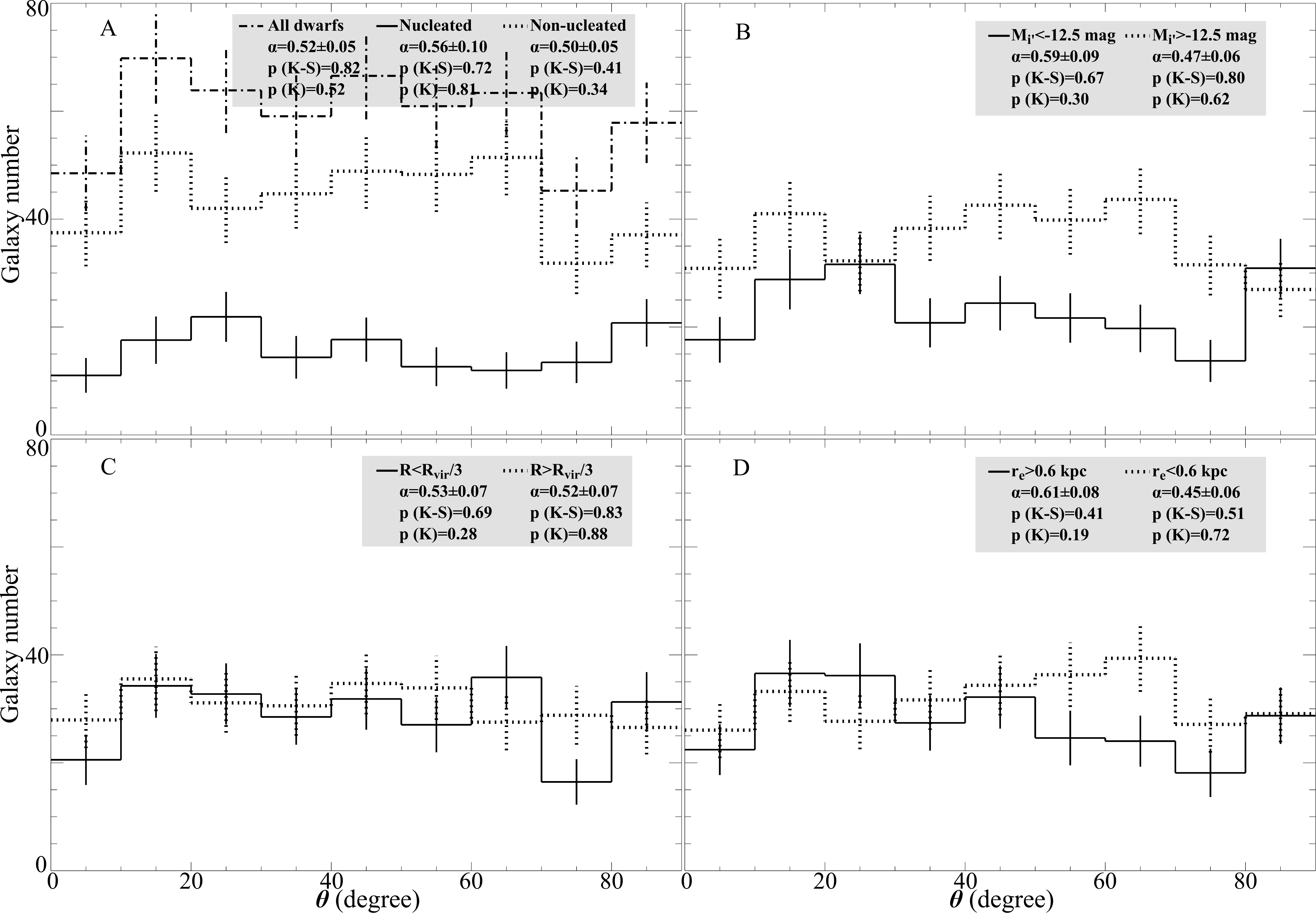}
\caption{\small Panel~A: the distributions of $\theta$ (PADs) for all of the Fornax dwarfs (dash-dotted), nucleated (solid), and non-nucleated (dotted) samples. Panel~B: PADs for the bright (solid) and faint (dotted) dwarfs. Panel~C: PADs for the inner-region (solid) and outer-region (dotted) dwarf samples. Panel~D: PADs for the large (solid) and small (dotted) dwarfs.}
\label{PA}
\end{figure*}

\subsection{Photometry}

In our previous NGFS papers \citep{Munoz15,Eigenthaler18,Yasna18a}, we 
reported the detection of the extended sources in the Fornax cluster from the RGB full-color image; in total, there are 643 dwarf galaxies which are very likely to be the members of the Fornax cluster (according to their locations in the color-magnitude diagrams, morphologies, and sizes; see Eigenthaler et al. 2018 for details), including the known 299 ones from previous works \citep{Ferguson89,Mieske07}. Among the 643 dwarfs, seven galaxies are ultra-diffuse galaxies \citep[UDGs;][]{vanDokkum15,Rong17}.
The sample include 181 nucleated and 462 non-nucleated dwarfs ranging the absolute $i'$-band magnitudes of $\sim -17.5\lesssim M_{i'}\lesssim -8.5$~mag. 

The two-dimensional surface brightness profile for each dwarf is studied with GALFIT (v3.0.5 Peng et al. 2002) implementing a S\'ersic profile, by taking advantage of the iterative fitting method described in Eigenthaler et al. (2018);
the best-fitting parameters, e.g., the magnitude, effective radius $r_{\rm{e}}$, position angle (PA) and its error PA\_error, etc., are obtained. PAs in the $g'$ and $i'$-bands coincide with each other very well; yet the $u'$-band PAs of several dwarfs significantly deviate from their $g'$ and $i'$-band PAs. Since the $g',i'$-bands are more likely to indicate the stellar mass distribution while $u'$-band is probably affected by the gas and current star-formation, we prefer to use the $i'$-band PAs to test the alignments. The radial angles $\phi$ and position angles $\theta$ of the dwarfs, used to quantitatively test the radial and primordial alignments, are then calculated from the $i'$-band PAs and locations of the dwarfs, as well as PA of BCG of Fornax, NGC~1399 \citep[with an axis ratio of $b/a\simeq 0.85$ and PA$\simeq 110^{\circ}$;][]{Schuberth10}.

\subsection{Radial alignment test}

We investigate the possible radial alignment signal for the Fornax dwarfs, abandoning the 94 ones with $b/a\geq0.9$ or large PA\_error$>10^{\circ}$. The distribution of the radial angles $\phi$ (RAD)
of the dwarfs is shown as the dash-dotted histogram in panel~A of Fig.~\ref{RA_final}.
We use the $p$ values returned from the Kolmogorov-Smirnov test (denoted as $p$ (K-S)) and Kuiper test (denoted as $p$ (K)) to detect the deviation of RAD from a uniform distribution; analogous to the work of \cite{Niederste-Ostholt10}, we also utilize the ratio $\alpha${\footnote{The uncertainties of $\alpha$ are estimated with the bootstrap methodology. For the galaxy population in each distribution, we resample the galaxies 10000 times, and obtain 10000 $\alpha$ values; the mean value and standard deviation of the 10000 values are treated as $\alpha$ and its uncertainty, respectively.}} between the numbers of galaxies with $\phi<30^{\circ}$ and $\phi>30^{\circ}$ to quantitatively assess the significance of the alignment signal (a uniform distribution corresponding to $\alpha \simeq 0.5$, while the radial alignment corresponding to $\alpha\gg0.5$). We find that, for the entire Fornax dwarf sample, $p\ \rm{(K-S)}\sim10^{-5}$, $p\ \rm{(K)}\sim10^{-6}$ and $\alpha\simeq 0.71\pm 0.06$,
suggesting a radial alignment of the Fornax dwarfs. We also test the alignment of the seven Fornax UDGs; the ratio $\alpha\sim 0.75\pm0.57$ indicates no significant radial alignment among the Fornax UDGs due to the large uncertainty.
This result is in consistent with the findings of \cite{Venhola17}, but different from that of UDGs in the Coma cluster \citep{Yagi16}, where the member UDGs are found to be tidally stretched towards BCG.



\cite{Sanchez-Janssen19} found that the morphologies of dwarf galaxies depend on the presence of NSCs and galaxy luminosities. Therefore, we divide the NGFS dwarfs into the nucleated and non-nucleated samples, and compare their RADs in panel~A of Fig.~\ref{RA_final}. The $\alpha$ and $p$ values for the different dwarf samples are also listed in Table~\ref{table} for comparison. The radial alignment signal of the nucleated dwarfs ($\alpha\simeq 0.96\pm 0.12$, $p\ \rm{(K-S)}\sim10^{-5}$, and $p\ \rm{(K)}\sim10^{-5}$) has higher significance than that of the non-nucleated dwarfs ($\alpha\simeq 0.63\pm 0.07$, $p\ \rm{(K-S)}\sim4\times10^{-2}$, and $p\ \rm{(K)}\sim5\times10^{-2}$) at $\sim 2.4\sigma$ level. The direct Kuiper test between the nucleated and non-nucleated samples also show a small $p\ \rm{(K)}\sim0.04$, excluding the radial angles of the two samples to follow the same distribution.

The radial alignment signals of the bright ($M_{i'}<-12.5$~mag) and faint ($M_{i'}>-12.5$~mag) dwarf samples are also compared, as shown in panel~B of Fig.~\ref{RA_final}. We obtain $\alpha\simeq 0.69\pm 0.10$, $p\ \rm{(K-S)}\sim5\times10^{-3}$, and $p\ \rm{(K)}\sim5\times10^{-3}$ for the bright ones, and $\alpha\simeq 0.73\pm 0.08$, $p\ \rm{(K-S)}\sim3\times10^{-4}$, and $p\ \rm{(K)}\sim2\times10^{-3}$ for the faint ones. Their $\alpha$ values which denote the significances of the radial alignment signals of the two samples differ at only $0.3\sigma$ level; further, the direct Kuiper test between the bright and faint samples gives $p\ \rm{(K)}\sim0.43$, suggesting that the significance of radial alignment signal is independent of the dwarf luminosities.

The dwarfs are also split into the inner-region sample with the projected distances to the cluster center of $R<R_{\rm{vir}}/3$ and outer-region sample with $R>R_{\rm{vir}}/3$, as well as the samples with large sizes and small sizes (with a threshold of the $i'$-band median $r_{\rm{e}}$ $\sim0.6$~kpc); their RADs, are shown in panels~C and D of Fig.~\ref{RA_final}, respectively. Panel~C reveals $\alpha\simeq 0.63\pm 0.09$, $p\ \rm{(K-S)}\sim2\times10^{-2}$, and $p\ \rm{(K)}\sim6\times10^{-2}$ for the inner-region sample, and $\alpha\simeq 0.81\pm 0.08$, $p\ \rm{(K-S)}\sim10^{-4}$, and $p\ \rm{(K)}\sim4\times10^{-4}$ for the outer-region sample, respectively, suggesting a slightly stronger radial alignment signal in the outer region at about $1.5\sigma$ level; the direct Kuiper test between the two samples also gives a relatively low $p\ \rm{(K)}\sim0.14$. In panel~D, the larger dwarfs exhibit $\alpha\simeq 0.73\pm 0.09$, $p\ \rm{(K-S)}\sim10^{-3}$, and $p\ \rm{(K)}\sim4\times10^{-3}$, while the smaller dwarfs show $\alpha\simeq 0.70\pm 0.09$, $p\ \rm{(K-S)}\sim5\times10^{-3}$, and $p\ \rm{(K)}\sim10^{-2}$, suggesting that the signals differ at only $0.2\sigma$ level; the direct Kuiper test between the large-size and small-size dwarf samples also reveal a high $p\ \rm{(K)}\sim0.68$. It indicates that the significance of radial alignment signal is also independent of the dwarf sizes.

\section{Discussion}\label{sec:3}

As clarified in Rong et al. (2015b), a fake radial alignment signal can be produced if the member galaxies exhibit primordial alignment ($\theta\to 0$), and simultaneously, are distributed along the major axis of BCG;
this fake signal may confuse us from identifying the real radial alignment of the Fornax dwarfs. Therefore, for the dwarf samples shown in the four panels of Fig.~\ref{RA_final}, we also plot their distributions of $\theta$ in the corresponding panels of Fig.~\ref{PA}. These dwarf samples show $\alpha\sim 0.5$, $p({\rm{K-S}})\gg 0$, and $p({\rm{K}})\gg 0$
suggesting no primordial alignment signal, regardless of the presence of NSCs, galactic locations, luminosities, or sizes; therefore, all of the detected radial alignment signals shown in Fig.~\ref{RA_final} are real.




\begin{figure}[!]
\centering
\includegraphics[width=\columnwidth]{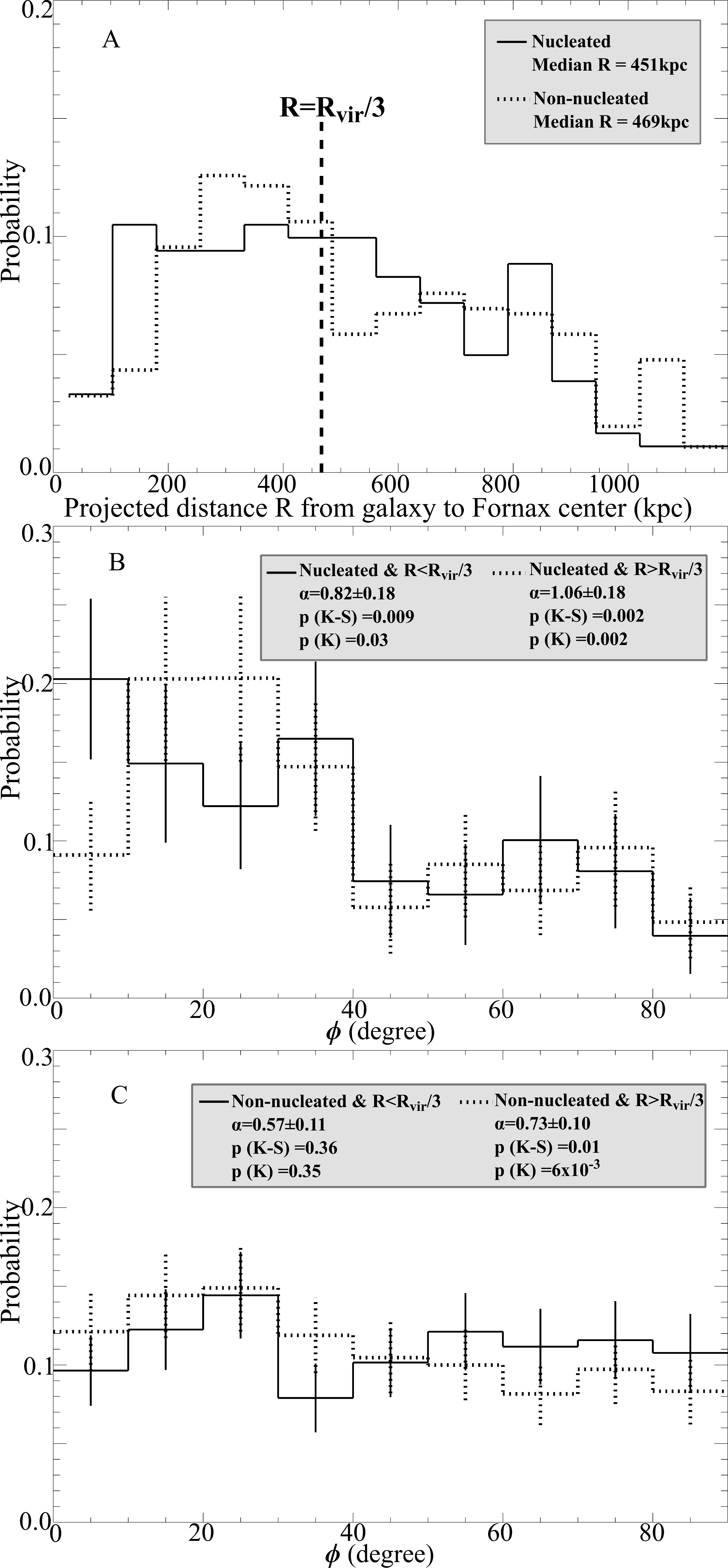}
\caption{\small Panel~A: The solid and dotted histograms denote the distributions of $R$ of the nucleated and non-nucleated Fornax dwarfs, respectively; the dashed line highlights the threshold of $R=R_{\rm{vir}}/3$; the median $R$ of the two samples are also listed in the panel. Panel~B: RADs of the nucleated dwarfs located at $R<R_{\rm{vir}}/3$ (solid) and $R>R_{\rm{vir}}/3$ (dotted). Panel~C: RADs of the non-nucleated dwarfs located at $R<R_{\rm{vir}}/3$ (solid) and $R>R_{\rm{vir}}/3$ (dotted).}
\label{dis}
\end{figure}

\begin{figure}[!]
\centering
\includegraphics[width=\columnwidth]{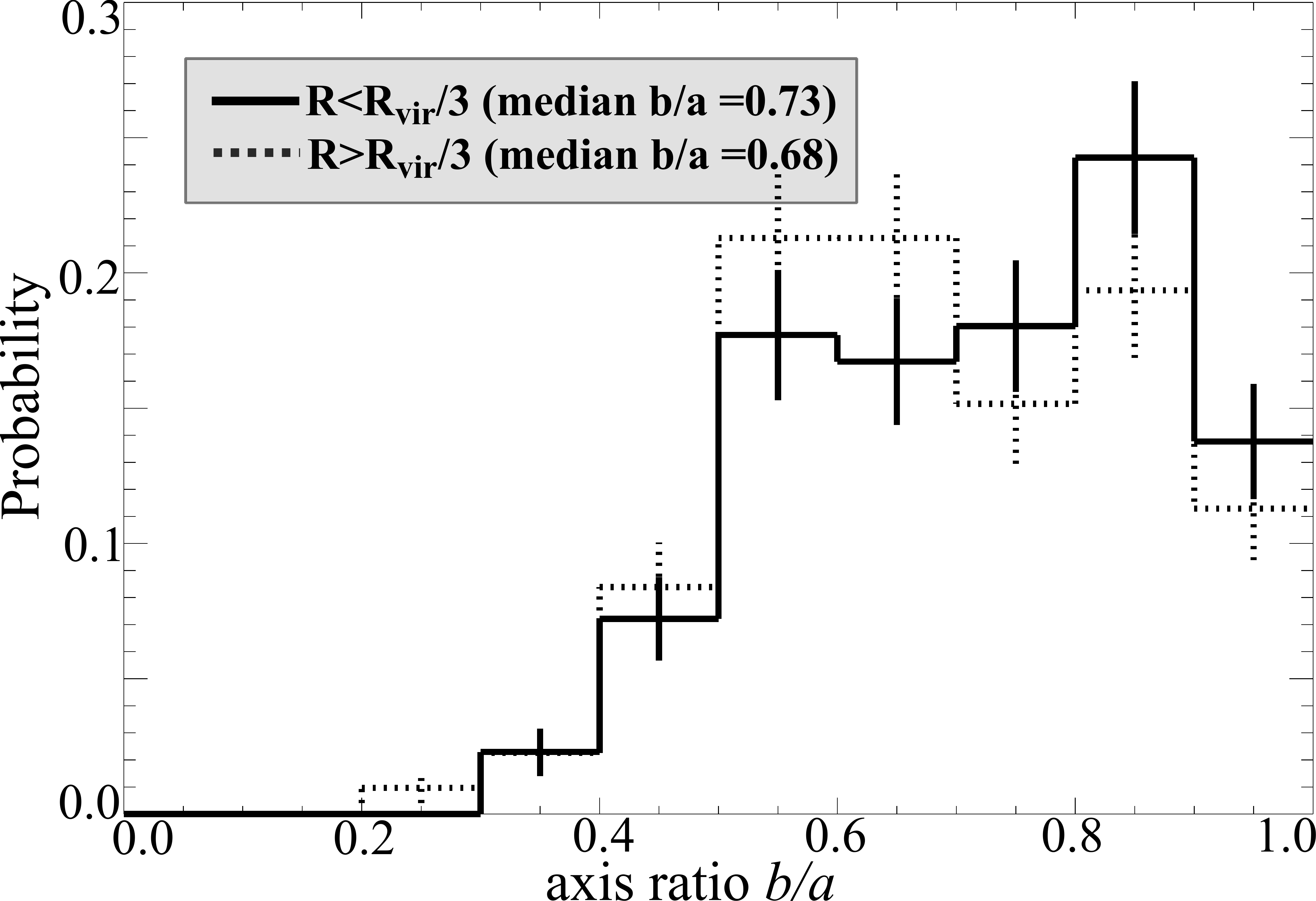}
\caption{\small The axis ratio ($b/a$) distributions for the dwarfs in the projected cluster-centric distance regions of $R<R_{\rm{vir}}/3$ (solid) and $R>R_{\rm{vir}}/3$ (dotted), respectively.}
\label{axisratio}
\end{figure}

The relatively remarkable radial alignment signals of the Fornax dwarfs conflict with the previous results of Barkhouse et al. (2016), in which the authors found no statistically significant radial alignment among the red-sequence cluster dwarfs over a redshift range of $0<z<0.35$. Barkhouse et al. (2016) also found no primordial alignment for their low-redshift red-sequence cluster dwarfs, which is, however, consistent with our results. We also find that changing the different $b/a$ cuts of dwarf samples from 0.9 to 0.85, 0.8, 0.75, and 0.7 does not change our conclusion.

The radial alignment signal is predominantly from the nucleated dwarfs, which may be due to the older stellar ages of the nucleated dwarfs as well as their earlier infall into the cluster, and thus longer time of stretching by the tidal force of Fornax (e.g., Rakos \& Schombert 2004).

Generally, the radial alignment signal of the inner-region galaxies should be stronger than that of the outer-region galaxies, since the tidal force in the inner region is stronger \citep[e.g.][]{Rong15a}. However in this work, the Fornax dwarfs located in $R>R_{\rm{vir}}/3$ show slightly stronger radial alignment signal than those located in $R<R_{\rm{vir}}/3$. 
It is not caused by the dominance of nucleated dwarfs in either sample, since they occupy similar fractions of 29\% and 31\% in the inner- and outer-region sample, respectively; the nucleated and non-nucleated dwarfs also follow the similar projected distributions, as explored in panel~A of Fig.~\ref{dis}. As shown in the panels~B and C of Fig.~\ref{dis}, actually both of the outer-region nucleated and non-nucleated samples exhibit stronger alignment signals, compared with the corresponding inner-region samples (at $\sim 1\sigma$ levels). Indeed, the mechanism leading to the weaker alignment signal of the inner-region dwarfs is still unclear. One plausible explanation is that the inner-region dwarfs move too fast at the pericenters and their deformation time-scales are longer than or comparable to the Keplerian time-scales, so that the inner-region dwarfs may not have enough time to align their major axes. As indicated by Rong et al. (2015a), the deformation time-scale of a galaxy is $\tau_{\rm{def}}\simeq 1/\sqrt{G\pi \rho_{\rm{G}}}$, where $G$ and $\rho_{\rm{G}}$ are the gravitational constant and mean galaxy density, respectively; whereas the Keplerian time-scale characterizing the rate of change of the tidal force is $\tau_{\rm{K}}\simeq 1/\sqrt{GM(L)/L^3}$, where $L$ and $M(L)$ are the three-dimensional (3D) distance from the galaxy to the cluster center and enclosed cluster mass within $L$, respectively. In order to estimate $\tau_{\rm{def}}/\tau_{\rm{K}}$, we assume a Navarro-Frenk-White profile $M(L)=4\pi\rho_0r_{\rm{s}}^3[{\rm{ln}}(1+L/r_{\rm{s}})-\frac{L/r_{\rm{s}}}{1+L/r_{\rm{s}}}]$ for the Fornax cluster with the scale radius $r_{\rm{s}}\simeq 150$~kpc, concentration $c\simeq 9$, and 
the virial mass of $\sim 10^{14}\ M_{\odot}$, as well as $\rho_{\rm{G}}\sim 10^7\ M_{\odot}/\rm{kpc}^3$ \citep{Dugger10,Drinkwater01,Oh15,Rong18}; therefore, for a innermost dwarf with $L\sim 100$~kpc, $\tau_{\rm{def}}/\tau_{\rm{K}}\sim 1$, suggesting that the rate of deforming the dwarf galaxy is comparable to the rate of change of the tidal force. 

The weaker radial alignment of the inner-region dwarfs may also attributed to projection effects, i.e., the so-called `inner-region' ($R<R_{\rm{vir}}/3$) dwarf sample is more likely to be dominated by the projected galaxies which are actually distributed in front and behind BCG with large 3D cluster-centric distances, compared with the outer-region sample. If we assume that all dwarfs are elongated towards BCG, the elongations of the front and back galaxies would lie roughly parallel to the line-of-sight; therefore, these projected dwarfs in the `inner-region' sample should tend to be round and show no significant radial alignment. As shown in Fig.~\ref{axisratio}, we compare the apparent axis ratio $b/a$ distributions of the `inner-region' and `outer-region' samples, and indeed find the marginally larger $b/a$ values for the `inner-region' sample. However, it is also worth to note that the rounder shapes of the inner-region dwarfs may also be produced by tidal interactions rather than projection effects \citep[e.g.,][]{Moore96,Mayer01,Mayer07,Rong19,Errani15}; the stronger tidal interactions in the inner regions of clusters (i.e., the denser environments) can efficiently puff up dwarf galaxies and also lead to the larger apparent axis ratios. Unfortunately, with the photometry data alone, we actually cannot distinguish the two possible rounding mechanisms.


In summary, we reported a significant radial alignment signal among the Fornax dwarfs with the NGFS data. For the first time, we found that the radial alignment signal depends on the presence of NSCs and locations of dwarfs in the cluster, but is independent of galaxy luminosities and sizes. The radial alignment signal of the nucleated dwarfs is stronger than that of non-nucleated ones at $2.4\sigma$ level, and the dwarfs located in the outer region ($R > R_{\rm{vir}}/3$) show slightly stronger radial alignment signal than those in the inner region ($R < R_{\rm{vir}}/3$) at $1.5\sigma$ level. Further alignment studies for the faint dwarfs in other massive clusters are required to test our findings.

\acknowledgments
Y.R. acknowledges supports through FONDECYT Postdoctoral Fellowship Project No.~3190354, NSFC grant No.\,11703037, and CAS-CONICYT Postdoctoral Fellowship Project CAS~16004. T.H.P. acknowledges support through FONDECYT Regular project 1161817 and CONICYT project Basal AFB-170002. M.A.T. is supported by the Gemini Observatory, which is operated by the Association of Universities for Research in Astronomy, Inc., on behalf of the international Gemini partnership of Argentina, Brazil, Canada, Chile, the Republic of Korea, and the United States of America. 
This work is also supported by CAS South America Center for Astronomy (CASSACA), Chinese Academy of Sciences (CAS).

This research has made use of the NASA Astrophysics Data System Bibliographic Services, the NASA Extragalactic Database, the SIMBAD database, operated at CDS, Strasbourg, France \citep{Wenger00}.~This research has made use of ``Aladin Sky Atlas'' \citep{Bonnarel00,Boch14}, developed at CDS, Strasbourg Observatory, France.~Software used in the analysis includes, the {\sc Python/NumPy} v.1.11.2 and {\sc Python/Scipy} v0.17.1 \citep[][\url{http://www.scipy.org/}]{Jones01,VanderPlas12}, {\sc Python/astropy} \citep[v1.2.1;][\url{http://www.astropy.org/}]{ast13}, {\sc Python/matplotlib} \citep[v2.0.0;][\url{http://matplotlib.org/}]{Hunter07}, {\sc Python/scikit-learn} \citep[v0.17.1;][\url{http://scikit-learn.org/stable/}]{Pedregosa12} packages.~This research made use of ds9, a tool for data visualization supported by the Chandra X-ray Science Center (CXC) and the High Energy Astrophysics Science Archive Center (HEASARC) with support from the JWST Mission office at the Space Telescope Science Institute for 3D visualization. We also acknowledge the related literatures of \cite{Press92,Paltani04,Rong17b,Rong18b,Johnston19}.\\







\begin{thebibliography}{}

\bibitem[\protect\citeauthoryear{Astropy Collaboration et al.}{2013}]{ast13} Astropy Collaboration, Robitaille, T. P., Tollerud, E. J., et al. 2013, A\&A, 558, A33
\bibitem[\protect\citeauthoryear{Barkhouse et al.}{2016}]{Barkhouse16} Barkhouse, W. A., Archer, H., Burgad, J., Foote, G., Rude, C., L\'opez-Cruz, O. 2016, Galaxies at High Redshift and Their Evolution Over Cosmic Time, IAU Symposium, Vol. 319, pp. 5-5
\bibitem[\protect\citeauthoryear{Bertin}{2011}]{Bertin11} Bertin, E. 2011, adass XX, 442, 435
\bibitem[\protect\citeauthoryear{Bertin \& Arnouts}{1996}]{ber96} Bertin, E., \& Arnouts, S.1996, \aaps,  117,  393
\bibitem[\protect\citeauthoryear{Boch \& Fernique}{2014}]{Boch14} Boch, T., \& Fernique, P. 2014, in ASP Conf. Ser. 485, Astronomical Data Analysis Software and Systems XXIII, ed. N. Manset \& P. Forshay (San Francisco, CA: ASP), 277
\bibitem[\protect\citeauthoryear{Bonnarel et al.}{2000}]{Bonnarel00} Bonnarel, F., Fernique, P., Bienaym\'e, O., et al. 2000, A\&AS, 143, 33
\bibitem[\protect\citeauthoryear{Ciotti \& Dutta}{1994}]{Ciotti94} Ciotti, L., Dutta S.~N. 1994, MNRAS, 270, 390
\bibitem[\protect\citeauthoryear{Drinkwater et al.}{2001}]{Drinkwater01} Drinkwater, M. J., Gregg, M. D., \& Colless, M. 2001, ApJL, 548, L139
\bibitem[\protect\citeauthoryear{Dugger et al.}{2010}]{Dugger10} Dugger, L., Jeltema, T. E., Profumo, S. 2010, Journal of Cosmology and Astroparticle Physics, 12, 15
\bibitem[\protect\citeauthoryear{Eigenthaler et al.}{2018}]{Eigenthaler18} Eigenthaler, P., Puzia, T.~H., Taylor, M.~A., et al.\ 2018, ApJ, 855, 142
\bibitem[\protect\citeauthoryear{Errani et al.}{2015}]{Errani15} Errani, R., Penarrubia, J., Tormen, G. 2015, MNRAS, 449, L46
\bibitem[\protect\citeauthoryear{Ferguson}{1989}]{Ferguson89} Ferguson, H. C. 1989, AJ, 98, 367
\bibitem[\protect\citeauthoryear{Flaugher et al.}{2015}]{Flaugher15} Flaugher, B., Diehl, H. T., Honscheid, K., et al. 2015, AJ, 150, 150
\bibitem[\protect\citeauthoryear{Fuller et al.}{1999}]{Fuller99} Fuller, T. M., West, M. J., Bridges, T. J. 1999, ApJ, 519, 22
\bibitem[\protect\citeauthoryear{Hirata \& Seljak}{2004}]{Hirata04} Hirata, C. M. \& Seljak, U. 2004, Phys. Rev. D, 70, 063526
\bibitem[\protect\citeauthoryear{Hung et al.}{2010}]{Hung10} Hung, L.-W., Ba{\~n}ados, E., De Propris, R., West, M.~J. 2010, ApJ, 720, 1483
\bibitem[\protect\citeauthoryear{Hunter}{2007}]{Hunter07} Hunter, J. D. 2007, CSE, 9, 90
\bibitem[\protect\citeauthoryear{Johnston et al.}{2019}]{Johnston19} Johnston, E. J., et al. 2019, ApJ, 873, 59
\bibitem[\protect\citeauthoryear{Jones et al.}{2001}]{Jones01} Jones, E., Oliphant, T., Peterson, P., et al. 2001, SciPy: Open Source Scientific Tools for Python, http://www.scipy.org/
\bibitem[\protect\citeauthoryear{Mayer et al.}{2001}]{Mayer01} Mayer, L., Governato, F., Colpi, M., Moore, B., Quinn, T., Wadsley, J., Stadel, J., Lake, G. 2001, ApJ, 547, L123
\bibitem[\protect\citeauthoryear{Mayer et al.}{2007}]{Mayer07} Mayer, L., Kazantzidis, S., Mastropietro, C., Wadsley, J. 2007, Nature, 445, 738
\bibitem[\protect\citeauthoryear{Mieske et al.}{2007}]{Mieske07} Mieske, S., Hilker, M., Infante, L., Mendes de Oliveira, C. 2007, A\&A, 463, 503
\bibitem[\protect\citeauthoryear{Moore et al.}{1996}]{Moore96} Moore, B., et al. 1996, Nature, 379, 613 
\bibitem[\protect\citeauthoryear{Mu{\~n}oz et al.}{2015}]{Munoz15} Mu{\~n}oz, R.~P., Eigenthaler, P., Puzia, T.~H., et al.\ 2015, \apjl, 813, L15 
\bibitem[\protect\citeauthoryear{Niederste-Ostholt et al.}{2010}]{Niederste-Ostholt10} Niederste-Ostholt, M., Strauss, M. A., Dong, F., Koester, B. P., McKay, T. A. 2010, MNRAS, 405, 2023
\bibitem[\protect\citeauthoryear{Oh et al.}{2015}]{Oh15} Oh, S.-H., Hunter, D. A., et al. 2015, ApJ, 149, 180
\bibitem[\protect\citeauthoryear{Ordenes-Brice{\~n}o et al.}{2018a}]{Yasna18a} Ordenes-Brice{\~n}o, Y., Eigenthaler, P., Taylor, M.~A., Puzia, T.~H., et al.\ 2018a, ApJ, 859, 52
\bibitem[\protect\citeauthoryear{Ordenes-Brice{\~n}o et al.}{2018b}]{Yasna18b} Ordenes-Brice{\~n}o, Y., Puzia, T.~H., Eigenthaler, P., et al. 2018b, ApJ, 860, 4
\bibitem[\protect\citeauthoryear{Paltani}{2004}]{Paltani04} Paltani, S. 2004, A\&A, 420, 789
\bibitem[\protect\citeauthoryear{Pedregosa et al.}{2012}]{Pedregosa12} Pedregosa, F., Varoquaux, G., Gramfort, A., et al. 2012, JMLR, 12, 2825
\bibitem[\protect\citeauthoryear{Peng et al.}{2002}]{Peng02} Peng, C.~Y., Ho, L.~C., Impey, C.~D., \& Rix, H.-W. 2002, \aj, 124, 266 
\bibitem[\protect\citeauthoryear{Pereira et al.}{2008}]{Pereira08} Pereira, M.~J., Bryan, G.~L., Gill, S.~P.~D. 2008, ApJ, 672, 825
\bibitem[\protect\citeauthoryear{Plionis et al.}{2003}]{Plionis03} Plionis, M., Benoist, C., Maurogordato, S., Ferrari, C., Basilakos, S. 2003, ApJ, 594, 144
\bibitem[\protect\citeauthoryear{Press et al.}{1992}]{Press92} Press, W. H., Teukolsky, S. A., Vetterling, W. T., Flannery, B. P. 1992, Numerical Recipes, 2nd edn. Cambridge Univ. Press, Cambridge
\bibitem[\protect\citeauthoryear{Rakos \& Schombert}{2004}]{Rakos04} Rakos, K. \& Schombert, J. 2004, AJ, 127, 1502
\bibitem[\protect\citeauthoryear{Rong et al.}{2019}]{Rong19} Rong, Y., Dong, X.-Y., et al. 2019, arXiv: 1907.10079
\bibitem[\protect\citeauthoryear{Rong et al.}{2015a}]{Rong15a} Rong, Y., Yi, S.-X., Zhang, S.-N., Tu, H. 2015a, MNRAS, 451, 2536
\bibitem[\protect\citeauthoryear{Rong et al.}{2015b}]{Rong15b} Rong, Y., Zhang, S.-N., Liao, J.-Y. 2015b, MNRAS, 453, 1577
\bibitem[\protect\citeauthoryear{Rong et al.}{2016}]{Rong16} Rong, Y., Liu, Y., Zhang, S.-N. 2016, MNRAS, 455, 2267
\bibitem[\protect\citeauthoryear{Rong et al.}{2017a}]{Rong17} Rong, Y., Guo, Q., Gao, L., Liao, S., Xie, L., Puzia, T. H., Sun, S., Pan, J. 2017a, MNRAS, 470, 4231
\bibitem[\protect\citeauthoryear{Rong et al.}{2017b}]{Rong17b} Rong, Y., Jing, Y., Gao, L., Guo, Q., Wang, J., Sun, S., Wang, L., Pan, J. 2017b, MNRAS, 471L, 36
\bibitem[\protect\citeauthoryear{Rong et al.}{2018a}]{Rong18} Rong, Y., Li, H., Wang, J., et al. 2018a, MNRAS, 477, 230
\bibitem[\protect\citeauthoryear{Rong et al.}{2018b}]{Rong18b} Rong, Y., et al. 2018b, arXiv: 1806.10149
\bibitem[\protect\citeauthoryear{Roychowdhury et al.}{2013}]{Roychowdhury13} Roychowdhury, S., Chengalur, J., Karachentsev, I., Kaisina, E. 2013, MNRAS, 436, L104
\bibitem[\protect\citeauthoryear{S\'anchez-Janssen et al.}{2019}]{Sanchez-Janssen19} S\'anchez-Janssen, R., Puzia, T., et al. 2019, arXiv: 1901.04509
\bibitem[\protect\citeauthoryear{S\'anchez-Janssen et al.}{2010}]{Sanchez-Janssen10} S\'anchez-Janssen, R., M\'endez-Abreu, J., Aguerri, J. A. L. 2010, MNRAS, 406, L65
\bibitem[\protect\citeauthoryear{Schuberth et al.}{2010}]{Schuberth10} Schuberth, Y., Richtler, T., et al. 2010, A\&A, 513, 52
\bibitem[\protect\citeauthoryear{Skrutskie et al.}{2006}]{Skrutskie06} Skrutskie, M. F., Cutri, R. M., Stiening, R., et al. 2006, AJ, 131, 1163
\bibitem[\protect\citeauthoryear{Struble}{1990}]{Struble90} Struble M. F., 1990, AJ, 99, 743
\bibitem[\protect\citeauthoryear{Sutherland et al.}{2015}]{Sutherland15} Sutherland, W., Emerson, J., Dalton, G., et al. 2015, A\&A, 575, A25
\bibitem[\protect\citeauthoryear{Troxel \& Ishak}{2015}]{Troxel14} Troxel, M. A. \& Ishak, M. 2015, Physics Reports, 558, 1
\bibitem[\protect\citeauthoryear{Umetsu et al.}{2014}]{Umetsu14} Umetsu, K., Medezinski, E., Nonino, M., et al. 2014, ApJ, 795, 163
\bibitem[\protect\citeauthoryear{van Dokkum et al.}{2015}]{vanDokkum15} van Dokkum, P. G., Abraham, R., Merritt, A., Zhang, J., Geha, M., Conroy, C. 2015, ApJ, 798L, 45
\bibitem[\protect\citeauthoryear{VanderPlas et al.}{2012}]{VanderPlas12} VanderPlas, J., Connolly, A. J., Ivezic, Z., \& Gray, A. 2012, in Proc. of Conf. on Intelligent Data Understanding (CIDU), 47
\bibitem[\protect\citeauthoryear{Venhola et al.}{2017}]{Venhola17} Venhola, A., et al. 2017, A\&A, 608, 142
\bibitem[\protect\citeauthoryear{Wenger et al.}{2000}]{Wenger00} Wenger, M., Ochsenbein, F., Egret, D., et al. 2000, A\&AS, 143, 9
\bibitem[\protect\citeauthoryear{West}{1994}]{West94} West, M. J. 1994, MNRAS, 268, 79
\bibitem[\protect\citeauthoryear{Yagi et al.}{2016}]{Yagi16} Yagi, M., Koda, J., Komiyama, Y., Yamanoi, H. 2016, ApJS, 225, 11

\end{thebibliography}
\end{document}